\documentclass[11pt,a4paper,twoside]{article}

\usepackage{epsfig}
\usepackage[latin1]{inputenc}
\def\be{\begin{equation}}
\def\ee{\end{equation}}
\def\ba{\begin{array}}
\def\bacc{\begin{array} {cc}}
\def\ea{\end{array}}
\def\bea{\begin{eqnarray}}
\def\eea{\end{eqnarray}}
\def\bd{\begin{displaymath}}
\def\ed{\end{displaymath}}

\begin{document}
\hspace{12cm} UAB-FT-678
\begin{center}

\vspace{1cm}

{\Large\bf Aspects of Brane Physics in 5 and 6 Dimensions}

\vspace{1cm}

{\large S. L. Parameswaran\footnote{
susha.parameswaran@fysast.uu.se, University of Uppsala}, S.
Randjbar-Daemi\footnote{seif@ictp.trieste.it, ICTP} and
A. Salvio\footnote{salvio@ifae.es, IFAE, Universitat Aut\`{o}noma de Barcelona}}

%\vspace{.6cm}

%{\it {$^a$ Scuola Internazionale Superiore di Studi Avanzati,\\
%Via Beirut 2-4, 34014 Trieste, Italy}}%
%
%\vspace{.4cm}%
%
%{\it {$^b$ International Center for Theoretical Physics, \\Strada
%Costiera 11, 34014 Trieste, Italy}}
%
%\vspace{.4cm}
%
%{\it {$^c$ Institut de Th\'eorie des Ph\'enom\`enes Physiques,
%\\EPFL, CH-1015 Lausanne, Switzerland}}

\end{center}

\vspace{1cm}

\begin{abstract}
We outline a general strategy to deal with perturbations in brane world models.  We illustrate the method by studying simple 6D brane world compactifications.  
We introduce the background geometries induced by the branes, and summarize the main physical features observed in the fluctuation spectra.  We compare these results with 5D brane world models and the traditional Kaluza-Klein picture.

\end{abstract}

\section{Introduction}

Extra dimensions, supersymmetry and branes have become a canonical ingredient in
physics beyond the standard model and cosmology.
In large part, this is due to their motivation from fundamental
theories of high energy physics, in particular string theory.  Indeed,
if string theory is to describe Nature, we must explain why only four
of the ten or eleven spacetime dimensions have been observed.

In the traditional Kaluza-Klein picture, a four dimensional effective field theory is obtained from a
higher dimensional one by compactifying the extra dimensions to a
very small size, say, the string scale.  Then, the higher dimensional
fields can be represented by an infinite tower of 4D Kaluza-Klein modes,
which are essentially harmonics in the internal manifold.  Massless
modes, such as the 4D graviton, are separated from the massive modes
by a finite mass gap.  This renders almost all the modes -- and thus the extra dimensions --
inaccessible to experiments performed at low energies.

The advent of branes widened our possibilities for hiding the extra
dimensions.  Some fields (the standard model) may be confined to a
brane in large extra
dimensions, in which case
we say their wave functions are localized on the brane.  Gravity, describing
the dynamics of spacetime itself, must of course propagate
everywhere.  However, an effective four dimensional gravitational
theory can be obtained if the massless mode of the
higher dimensional graviton is exponentially enhanced at our brane,
whereas the massive modes are exponentially suppressed.  This was
shown to happen in the 5D brane world model of Randall-Sundrum
\cite{Randall:1999ee}.

5D brane world models have been extremely well studied.  They are quite special, since the branes represent boundaries in the spacetime.  Much of the
interesting dynamics is due to the warp factor, which appears when
solving the gravitational backreaction of the brane.  It is
exponential in the proper distance from the brane.  This gives rise to
qualitatively different physics from the traditional Kaluza-Klein
picture, such as the possibility of infinitely large extra dimensions.
The Randall-Sundrum toy models have by now been realized in more
stringy scenarios, and are playing a role in such issues as the moduli
stabilization problem, inflation and AdS/CFT.

Indeed, extra dimensions are not only obstacles in our quest to
connect string theory to Nature.  Together with branes and
supersymmetry, they also
provide new avenues in which to approach some of the long-standing
problems of the standard model of particle physics and cosmology.  For
instance, the
exponential warping induced by codimension one branes, was proposed as
a solution to the electroweak hierarchy problem \cite{Randall:1999ee}.
Two submillimeter sized
extra dimensions were also proposed early on as a way to relate the
Planck and electroweak scales \cite{LED}, and together with supersymmetry, may
even provide a link between that hierarchy and the cosmological
constant problem \cite{SLED}.

In the following, we review some of our recent work exploring 6D brane
models \cite{Parameswaran:2006db,Parameswaran:2007cb,Parameswaran:2009bt}.  As for 5D, and unlike higher codimensions, the gravitational
backreaction of codimension two branes can be solved.  In the 5D case, the branes induce step-like singularities in the geometry, whereas in 6D the singularities are conical.  The warping, rather than the
exponential behaviour in 5D, is only power law.  We present various
background solutions induced by codimension two branes, and study
the spectra of small fluctuations near the backgrounds.  This analysis
is necessary to establish the stability of the configurations, and
also allows one to infer properties of the low energy effective 4D
theory, and when and how extra-dimensional physics comes into play.

In the next section, we describe the gravitational backreaction of
codimension two branes, and present the configurations to be
studied.  In Section 3 we outline the strategy for analysing the
perturbations, which can actually be applied to more general dimensions and
codimensions, and in Section 4 we summarize the main physical
results for 6D.  Finally, we conclude, by comparing and contrasting 6D brane
models with their 5D cousins and the traditional Kaluza-Klein models.

We end this introductory section with a few historical remarks. 
The idea of a brane world and matter localization on them is rather old and has been discussed in different contexts 
and with different motivations for more than two decades. In \cite{Jackiw:1981ee} the localization of fermion zero modes
on a U(1) vortex was discussed and  similar ideas were applied to the discussion of superconducting 
cosmic strings in \cite{Witten:1984eb}. The localization of massless gauge fields on a Euclidean 4-brane have even be applied to 
construct models of chiral fermions on the lattice \cite{Narayanan:1994gw}, and has been extended to include gravitational fields \cite{RandjbarDaemi:1994tj}. 
The application to particle physics and cosmology starts with the papers in \cite{Rubakov:1983bb}.

\newpage

\section{Codimension Two Brane Models}
\begin{figure}
\centering \epsfig{file=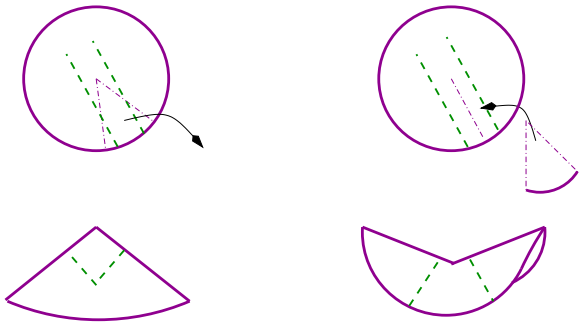,width=0.5\linewidth,clip=}
\caption{\small Construction of a cone and saddle-cone by splicing out and in, respectively, a wedge starting
from the flat disk. Reproduced from \cite{Parameswaran:2007cb}.  For the cone, lines that were parallel on the
disk remain parallel until they pass on either side of the apex,
when they begin to converge.  For the saddle-cone they diverge once
passing the apex.  Notice that although the saddle-cone appears to
break the axial-symmetry, this is only an effect of the embedding
into 3D.  A 2D being would indeed observe the axial
symmetry.}\label{fig:geodesics}
\end{figure}
A brane in higher dimensional space-time can be introduced with a
 function\footnote{We denote higher dimensional coordinates with
 $X^M$, and the world volume coordinates on the brane with
 $x^{\alpha}$.} $Y^M(x)$, which represents the position of the brane
 \cite{Sundrum:1998sj}.  With this function and the bulk metric,
 $G_{MN}$, we build the induced metric on the brane, $g_{\alpha\beta}=
 G_{MN}(Y(x)) \partial_{\alpha} Y^M \partial_{\beta}Y^N$, and we can
 then write an effective brane action of the form:
\be S_{brane}=\int d^4x\sqrt{-g}(-T+...),\label{Sbrane}\ee
 where $T$ is a constant parameter (the tension of the brane) and the
dots represent a series of higher derivative terms involving the
brane metric and any extra brane localized
fields. The total action is then
\be S_{bulk}+S_{brane}. \label{totalS}\ee
A famous example of this type is the Randall-Sundrum model
\cite{Randall:1999ee} in which $S_{bulk}$ is the 5D Einstein-Hilbert
action with cosmological constant, and one or two branes are included.

Let us discuss in some more detail the 6D case. Henceforth, we
assume that any brane localized matter is integrated out, leaving
pure tension branes.  The simplest choice
for the bulk action is the 6D Einstein-Hilbert term
\be S_{EH}=\frac{1}{\kappa^2}\int d^6 X \sqrt{-G}R,\label{Einstein}\ee
where $G$ is the determinant of the higher dimensional metric, and $\kappa$ the 6D Planck scale. For $S_{bulk}=S_{EH}$, a solution of
(\ref{totalS}) with 4D Poincar\'e invariance is given by the product
of the 4D Minkowski space-time and a cone
\be ds^2=\eta_{\mu \nu}dx^{\mu}dx^{\nu} +dr^2 +
\left(1-\frac{\delta}{2\pi}\right)^2 r^2 d \varphi^2.
\label{cone}\ee
The cone, whose metric is given above in polar coordinates, can be
constructed by splicing out a wedge starting from a flat space (see the example on the left in Figure \ref{fig:geodesics}). The
size of the deficit angle, $\delta$,
is related to the brane physics through
\be \delta=\frac{\kappa^2 T}{2}.\label{deltaT}\ee
The space remains flat apart from
at the apex, where the Ricci scalar acquires a delta-function behaviour.
 Since the deficit angle is bounded from above
by $2\pi$, we have an upper bound on the brane tensions that can be
described with these mild conical singularities. At the same time, the
solution in (\ref{cone}) is valid also when $\delta<0$.   Negative deficit angles are perfectly well
defined, and can even be made at home with a piece of paper and a
pair of scissors. A cone with negative deficit angle, which we call
saddle-cone, is obtained by
splicing a wedge {\it into} a flat space (see the example on the right
in Figure
\ref{fig:geodesics}).  It is
clear that the
deficit angle can even take arbitrarily large negative
values.   Meanwhile, the instabilities generically
associated with negative tension branes may be avoided by placing them
on orbifold fixed points, which projects out the unstable ``branon'' modes.

%\subsection{Geometry and topology} \label{S:geometry}

A simple generalization of the bulk action (\ref{Einstein}) is obtained by adding a
6D cosmological constant $\Lambda$ and Yang-Mills gauge fields
$\mathcal{A}$, with field strength $F$:
\be S_{bulk} =\int d^6 X \sqrt{-G}\left(\frac{1}{\kappa^2}R -\Lambda
-\frac{1}{4}F^2\right).\label{Sbulk2}\ee
This allows for the simplest models of flux compactifications
(relevant in moduli stabilization),
an example of which is given by the following ``rugby-ball'' solution,
in spherical-polar coordinates
\cite{Carroll:2003db}:
\be ds^2=\eta_{\mu
\nu}dx^{\mu}dx^{\nu}+\mathcal{R}^2\left(d\theta^2+\alpha^2\sin^2\theta
d\varphi^2\right), \quad \mathcal{A}=\frac{cQ}{2}(\cos\theta \pm 1
)d\varphi \equiv A_{\pm}.\label{sol-rugby-ball}\ee
Here, $Q$ is a generator of an Abelian subgroup of a
simple factor of the gauge group\footnote{We take the convention
$\mbox{Tr}\left(Q^2\right) =1$}.

In order for (\ref{sol-rugby-ball}) to be a solution to the equations of motion the
constants $\mathcal{R}$, $\alpha$ and $c$ have to fulfil  $c=
2\sqrt2 \mathcal{R} \alpha/\kappa$, $\mathcal{R}^2 =1/(\kappa^2
\Lambda)$. Thus we see that the size of the extra dimensions is fixed.
The metric has two conical singularities at $\theta=0$ and
$\theta=\pi$, with deficit angles $\delta=2\pi\left(1-|\alpha|
\right)$, which are sourced by two branes of equal tension\footnote{This fine-tuning,
however, can be avoided by considering more general solutions, e.g.
involving a warp factor.} as in Eq. (\ref{deltaT}).
 \begin{figure}
\centering \epsfig{file=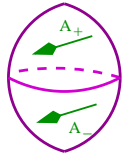,width=0.1\linewidth,clip=}
\caption{\small The rugby-ball. Two patches are needed to describe
this geometry. The representations of the gauge field in the two
patches, $A_+$ and $A_-$, have to be related by a single valued
gauge transformation, which leads to the Dirac quantization
condition.}\label{rugby-ball}
\end{figure}
 Finally, topological arguments, which are explained in Figure \ref{rugby-ball} and its caption, lead
to the so called Dirac quantization condition, which for a field
interacting with $\mathcal{A}$ through a charge $e$ gives
\be \overline{g} e c \equiv N =\mbox{integer},\ee
where $\overline{g}$ is the gauge coupling constant of the subgroup generated
by $Q$. The limit $\alpha\rightarrow 1$ (which is equivalent to the
limit of zero deficit angle or zero tension) corresponds to a
smooth flux  compactification on a sphere with radius $\mathcal{R}$ \cite{nonsusysphere}.

As motived in the introduction, it is also interesting to consider
more sophisticated models, with local supersymmetry in the bulk.  6D
chiral gauged supergravity \cite{Nishino:1984gk}, with its positive
definite scalar potential, provides a
supersymmetric version of the
Einstein-Yang-Mills systems discussed above. The bosonic sector
includes, additionally, a dilaton $\phi$ and a 2-form field $B_{MN}$,
which emerge from the
graviton multiplet and an antisymmetric tensor multiplet
\cite{Nishino:1984gk}. Besides the fermionic fields needed to
supersymmetrize this bosonic field content, one can also add  a number
of hypermultiplets
($\Phi$, $\Psi$) - with hyperscalars and hyperinos respectively. The
bulk action (for vanishing hyperscalars) is
then
\bea  S_{bulk}=&&\int d^6X
\sqrt{-G}\left\{\frac{1}{\kappa^2}\left[R-\frac{1}{4}\left(\partial
      \phi\right)^2\right]
-\frac{1}{4}e^{\phi/2}F^2
-\frac{\kappa^2}{48}e^{\phi}H_{MNP} H^{MNP} -\frac{8 g_1^2}{\kappa^4} e^{-\phi/2}\right\}\nonumber \\&& +\rm{ fermion \,\, terms},
\label{SB} \eea
% \ee
%
where
 \be H_{MNP} = \partial_M B_{NP} +
F_{MN}{\cal A}_P - \frac{g}{3}{\cal A}_M\left({\cal A}_N
\times {\cal A}_P \right) + \,\rm{2 \,\, cyclic \,\, perms} \,,\label{HKR}\ee
and $g_1$ is the gauge coupling of the $U(1)_R$ gauged R-symmetry.

The chiral fermions present in the theory generically lead to
gravitational and gauge anomalies, which can sometimes be cancelled by
the Green-Schwarz mechanism \cite{seifanomaly}.  This places an
attractive restriction on the matter content of the model, and
in Table \ref{anomalyfree} we give some examples of the anomaly free
models that have been discovered.

\begin{table}[top]
\begin{center}
\begin{tabular}{|l|l|}
\hline  Gauge Group & Hyperino Representation  \\ \hline
 $E_7 \times E_6 \times U(1)_R$ & $({\bf 912},{\bf
1})_0$ \\
 $E_7 \times G_2 \times U(1)_R$ & $({\bf 56},{\bf 14})_0$  \\
$F_4 \times Sp(9) \times U(1)_R$ & $({\bf 52},{\bf 18})_0$
 \\ \hline
\end{tabular}
\end{center}\caption{\footnotesize Some examples of anomaly free models with gauge groups containing $SU(3)\times SU(2) \times U(1)$ \cite{seifanomaly,Avramis:2005qt,Avramis:2005hc}.
 There are also many other models \cite{Suzuki:2005vu}. \label{anomalyfree}}
\end{table}

String theory derivations of 6D chiral supergravity have been provided \cite{Cvetic:2003xr} and its vacuum structure has been investigated in great detail.
Gibbons, G\"uven and Pope (GGP) proved \cite{Gibbons:2003di} that the
only smooth solution of the minimal (anomalous) model, with maximal
symmetry in 4D, is the unwarped 4D Minkowski spacetime, with an internal spherical
geometry supported by a non vanishing gauge flux \cite{salamsezgin}.
As soon as 3-branes are introduced, conical singularities are
generated and the most general solution with 4D maximal symmetry,
axial symmetry in the
internal space and only those mild conical singularities, has been
derived \cite{Gibbons:2003di,Aghababaie:2003ar}.  We consider this
class of solutions here and refer to it as the GGP solution.  It has
the form (notice the flat 4D slices; for an explicit expression see e.g. \cite{Parameswaran:2009bt}).
\bea
ds^2&=&e^{A(\theta)}\left(\eta_{\mu
\nu}dx^{\mu}dx^{\nu}+\mathcal{R}^2d\theta^2\right)+C(\theta)d\varphi^2 \nonumber\\
\mathcal{A}&=& \mathcal{A}_{\varphi}(\theta) d\varphi,\quad  \phi =\phi(\theta),\quad
H = 0,\nonumber\eea
Without warping ($A= 0$) the
internal space is a rugby ball. The warping deforms this solution, and
allows brane sources of different tensions.  However, the topology of
the internal space is still that of the sphere
and so we have a Dirac quantization condition, which relates the brane
tensions via a topological relation.  Almost all the solutions
spontaneously break the bulk supersymmetry.  The supersymmetric
configurations are unwarped, with the monopole flux embedded in the
$U(1)_R$ gauged R-symmetry, monopole number $N=1$, and brane tensions
such that $\delta= 0, -2\pi, -4\pi, \dots$.

\section{General Perturbations}

We now examine the small perturbations about the brane world
solutions. For simplicity, in this review we focus on the non-susy 6D model
(\ref{Sbulk2}) and its rugby ball solution, but the scheme
presented can be applied to other situations, such as the
supersymmetric models described above
\cite{Parameswaran:2006db,Parameswaran:2007cb,Parameswaran:2009bt} (see also \cite{kicking,Salvio:2007mb, gravitino, Salvio:2009mp}), the 5D Randall-Sundrum models or indeed smooth Kaluza-Klein compactifications.

Perturbing the solution
\be  G_{MN}\rightarrow G_{MN}+h_{MN},  \,\, \mathcal{A}_M\rightarrow
\mathcal{A}_M +V_M, \,\, \mbox{and} \,\, Y^M \rightarrow Y^M +\xi^M\nonumber \ee
one obtains a bilinear action for the perturbations $h_{MN}, V_M$ and $\xi^M$, which has local symmetries
descending from the 6D general covariance, the 6D gauge symmetry and
the brane general covariance. We fix the first two symmetries by
imposing the light cone gauge condition\footnote{The $\pm$
components of a generic vector $\mathcal{V}^M$ are defined by
$\mathcal{V}^{\pm} \equiv \left(\mathcal{V}^3\pm
\mathcal{V}^0\right)/\sqrt{2}$.}
\be h_{-M}=0,\,\, V_{-}=0.\ee
We fix instead the latter symmetry by requiring the static gauge
(that is $\xi^{\mu}=0$). In these gauges, perturbations with different
helicity decouple \cite{seifgen}, and one can examine separately
massless and massive gravitons \cite{Parameswaran:2009bt}, vectors and gauge invariance \cite{Parameswaran:2006db,Parameswaran:2009bt},
scalars \cite{Parameswaran:2007cb,Parameswaran:2009bt} and fermions \cite{Parameswaran:2006db}.

The graviton sector is particularly simple as the corresponding
bilinear action is proportional to
\be \int d^6X \sqrt{-G}\, \partial_M h
\partial^M h, \ee
where we have understood the Lorentz indices on the perturbations.
The corresponding equations of motion are
\be  \partial_M \left( \sqrt{-G} \,\partial^M h\right)=0.   \ee
In deriving this equation we have demanded that a boundary term
vanishes; this leads to boundary conditions \cite{Nicolai:1984jg,Gibbons:1986wg}. The system can then be translated into an equivalent Schroedinger
problem, which can be solved explicitly despite the presence of conical singularities.  The boundary conditions,
together with the requirement of finite kinetic energy for the
fluctuations, lead to a
discrete mass spectrum:
\be M^2 = \frac{1}{\mathcal{R}^2}({\bf n}+|{\bf m}|\omega)({\bf n}+|{\bf
m}|\omega +1),\nonumber  \ee
 where $\omega \equiv 1/\alpha$, and ${\bf
n}=0,1,2,...$ and ${\bf m}=0, \pm 1, \pm 2$ are the Kaluza-Klein wave-numbers. The wave functions can
be expressed in terms of the hypergeometric functions (see \cite{Parameswaran:2006db,Parameswaran:2009bt} for details).  These results can be used to compute the
 interactions mediated by massive gravitons on the brane \cite{Salvio:2009mp}.

A similar technique can be applied to analyze the fields with
different spin.  In the supersymmetric models, we encounter apparently
formidable systems of coupled 2nd order differential systems.  The
smaller systems can be solved in the full generality of the warped
models, and even the large ones can be tackled in the unwarped case,
with the development of so-called ''rugby ball harmonics''.

\section{Results}

We now present the main features of 6D brane worlds revealed by
our analysis of their perturbations.  It is interesting to compare and
contrast their behaviour with standard Kaluza-Klein compactifications
and 5D brane worlds.

\paragraph{Zero modes}

The standard lore of Kaluza-Klein theory (with or without branes) is
that any zero modes in the bosonic perturbation spectrum can be identified
due to the symmetries of the theory and background solution\footnote{Similarly for the gravitino in supersymmetric Kaluza-Klein models.  The gravitino spectrum in 6D brane world models was instead studied in \cite{gravitino}.}.  This is
indeed the case in 6D models with positive tension branes or with a smooth internal space (see e.g. \cite{RandjbarDaemi:2006gf}).  General
coordinate invariance in 4D gives rise to a zero mode for the metric, the
axial isometry in 2D gives rise to a massless gauge field, and the
classical scaling symmetry, $G_{MN} \rightarrow \lambda \, G_{MN}$ and
$e^{\phi} \rightarrow \lambda \, e^{\phi}$, and two-form gauge
symmetry, $B\rightarrow B+d\Omega$, each give rise to a massless
scalar.

In contrast, in particular models with negative tension branes, extra
massless vector fields appear in the spectrum, due to the presence of {\it infinitesimal}
isometries in the internal space.  The massless vector fields
are gauge fields in the low energy 4D effective theory obtained by truncating the massive modes.  However, the massive modes do not fall into well-defined representations of the corresponding gauge group, so it is not a genuine Kaluza-Klein gauge symmetry group.  This can be
understood as not all the infinitesimal isometries
can be integrated to genuine isometries.  Therefore, we expect the extra massless vector fields
to remain massless only at the classical level.

Meanwhile, for the fermionic zero modes, it is well known that massless chiral fermions do not generically arise in smooth Kaluza-Klein compactifications.  One way to obtain them is by turning on a non-trivial topological configuration for a gauge field background, such as a monopole flux \cite{nonsusysphere}.  Negative decifit angles in 6D brane models provide another mechanism for making massless fermions \cite{Schwindt:2003er}.  At the same time, the orbifold projections that feature in the 5D Randall-Sundrum models and models with negative tension branes also lead to chirality.

\paragraph{Mass gap}

The compactness of the internal spaces studied ensure that any zero
modes are separated from the rest of the spectrum by a mass gap.  In
standard Kaluza-Klein theory (and 5D brane worlds) as the volume of
the extra dimensions is taken to infinity, and mass gap goes to zero.
Again, this is also what is observed in 6D models with positive
tension branes.

The presence of negative tensions allows for another possibility.  If
the volume is made large by taking large negative deficit angles, the
mass gap can remain large too!  In principle, this allows for the
possibility of even the Standard Model fields to propagate in large
extra dimensions.

\paragraph{Wave function localization}

In 5D brane worlds, it is possible that the heavy Kaluza-Klein modes
are hidden due to their wave function being suppressed at the position
of the brane, where zero modes are instead exponentially enhanced.  In
6D, there is a universal asymptotic behaviour for all modes in a given
tower.  This can be inferred from the potential in the
effective Schroedinger problem, in which all dependence on the
Kaluza-Klein wave number is dominated at the boundaries by divergences
due to the conical defects.

\paragraph{Stability}

For the brane configurations discussed to be of interest, they should
be stable to small perturbations.  Indeed, most of the sectors are
free of tachyonic or ghost instabilities.  The only sector in which
instabilities may lurk are the scalar fluctuations, $V_m$, descending from the
6D gauge fields and charged under the background monopole.

\begin{table}[top]
\begin{center}
\begin{tabular}{|l|l|l|l|}
\hline & Zero Modes & Mass Gap & Wave Function \\ \hline
Kaluza-Klein & {\it due to symmetries} & {\it goes to zero with} &
{\it massless and massive}\\
& & {\it infinite volume} &  {\it modes have} \\
& & & {\it same behaviour} \\ \hline
5D Brane Worlds & {\it due to symmetries} & {\it goes to zero with} &
{\it massless modes may} \\
& & {\it infinite volume} & {\it be localized on} \\
& & &  {\it different brane} \\
& & & {\it to massive modes} \\ \hline
6D Brane Worlds &  {\it due to symmetries} &  {\it goes to zero with}
& {\it massless and massive} \\
($T >0$) & & {\it infinite volume} &  {\it modes have}  \\
& & &  {\it same behaviour} \\  \hline
6D Brane Worlds & {\it massless modes may} &  {\it can remain finite}
& {\it massless and massive} \\
($T < 0$) & {\it also arise due to} &  {\it with infinite volume} & {\it
  modes have}  \\
& {\it infinitesimal isometries} & & {\it same behaviour} \\ \hline
\end{tabular}
\end{center}\caption{\footnotesize Comparison of traditional
 Kaluza-Klein models, 5D Randall-Sundrum like models, and 6D models
 with positive tension branes only and with negative tension
 branes. In 5D Randall-Sundrum models, an effective 4D theory may be
 obtained even with infinitely large extra dimensions, due to the wave function
 localization properties of the graviton spectrum.  In contrast, with negative
 tension branes, all
 the Kaluza-Klein massive modes can be hidden in 6D models, due to
 the mass gap. \label{summary}}
\end{table}

The source of the instability is the monopole flux in a non-Abelian
gauge sector, and it is to be
found also in smooth sphere compactifications
\cite{seifinstabilities}.  It turns out that models that are stable
for the sphere, are also stable in the presence of positive tension
branes.  This includes all models in which the monopole is embedded in
an Abelian gauge group, even if they are non-supersymmetric.
Curiously, the introduction of negative tensions can render
an unstable monopole flux stable.

Possible endpoints of these instabilities have been studied in
\cite{endpoint}.

\section{Conclusions}

We have reviewed the physics of 6D brane worlds, focusing on the
background geometries that they induce and the behaviour of their
perturbations.  In the series of papers
\cite{Parameswaran:2006db,Parameswaran:2007cb,Parameswaran:2009bt} we
developed a fairly comprehensive analysis of the perturbations in
(warped) brane world compactifications of 6D chiral supergravity.  The
techniques developed could also applied to other scenarios, with
different dimensions and codimensions.

Our results show that the behaviour of bulk fluctuations in 6D brane
worlds, with positive tensions, is much the same as in traditional
Kaluza-Klein compactifications.  Moreover, the power law warping that
appears in 6D changes the physics little with respect to unwarped
models.  Instead, models with negative tension branes can lead to new physics.  In Table \ref{summary} we provide a summary, which compares
and contrasts the behaviour amongst Kaluza-Klein models, 5D
Randall-Sundrum like models and the 6D models discussed here.

For details, we encourage the reader to refer to \cite{Parameswaran:2009bt}.

\vspace{0.3cm}

{\bf Acknowledgments.} The work of A. S. has been
supported by  CICYT-FEDER-FPA2008-01430.  S. L. P. is supported by the G\"{o}ran Gustafsson Foundation.


\begin{thebibliography}{99}





%\cite{Randall:1999ee}
\bibitem{Randall:1999ee}
  L.~Randall and R.~Sundrum,
  %``A large mass hierarchy from a small extra dimension,''
  Phys.\ Rev.\ Lett.\  {\bf 83} (1999) 3370
  [arXiv:hep-ph/9905221].
  %%CITATION = PRLTA,83,3370;%%
 L.~Randall and R.~Sundrum, Phys.\ Rev.\ Lett.\  {\bf 83} (1999) 4690
  [arXiv:hep-th/9906064].
  %%CITATION = PRLTA,83,4690;%%

%\cite{ArkaniHamed:1998rs}
\bibitem{LED}
  N.~Arkani-Hamed, S.~Dimopoulos and G.~R.~Dvali,
  %``The hierarchy problem and new dimensions at a millimeter,''
  Phys.\ Lett.\  B {\bf 429} (1998) 263
  [arXiv:hep-ph/9803315].
  %%CITATION = PHLTA,B429,263;%%
%\cite{Antoniadis:1998ig}
%\bibitem{Antoniadis:1998ig}
  I.~Antoniadis, N.~Arkani-Hamed, S.~Dimopoulos and G.~R.~Dvali,
  %``New dimensions at a millimeter to a Fermi and superstrings at a TeV,''
  Phys.\ Lett.\  B {\bf 436} (1998) 257
  [arXiv:hep-ph/9804398].
  %%CITATION = PHLTA,B436,257;%%

%\cite{Aghababaie:2003wz}
\bibitem{SLED}
  Y.~Aghababaie, C.~P.~Burgess, S.~L.~Parameswaran and F.~Quevedo,
  %``Towards a naturally small cosmological constant from branes in 6D
  %supergravity,''
  Nucl.\ Phys.\  B {\bf 680} (2004) 389
  [arXiv:hep-th/0304256].
  %%CITATION = NUPHA,B680,389;%%
%\cite{Burgess:2004ib}
%\bibitem{Burgess:2004ib}
  C.~P.~Burgess,
  %``Towards a natural theory of dark energy: Supersymmetric large extra
  %dimensions,''
  AIP Conf.\ Proc.\  {\bf 743} (2005) 417
  [arXiv:hep-th/0411140].
  %%CITATION = APCPC,743,417;%%

\bibitem{Parameswaran:2006db}
  S.~L.~Parameswaran, S.~Randjbar-Daemi and A.~Salvio,
  %``Gauge fields, fermions and mass gaps in 6D brane worlds,''
  Nucl.\ Phys.\  B {\bf 767}, 54 (2007)
  [arXiv:hep-th/0608074].
  %%CITATION = NUPHA,B767,54;%%

%\cite{Parameswaran:2007cb}
\bibitem{Parameswaran:2007cb}
  S.~L.~Parameswaran, S.~Randjbar-Daemi and A.~Salvio,
  %``Stability and Negative Tensions in 6D Brane Worlds,''
  JHEP {\bf 0801} (2008) 051
  [arXiv:0706.1893 [hep-th]].
  %%CITATION = JHEPA,0801,051;%%
%\cite{Parameswaran:2009bt}
\bibitem{Parameswaran:2009bt}
  S.~L.~Parameswaran, S.~Randjbar-Daemi and A.~Salvio,
  %``General Perturbations for Braneworld Compactifications and the Six
  %Dimensional Case,''
  JHEP {\bf 0903} (2009) 136
  [arXiv:0902.0375 [hep-th]].
  %%CITATION = JHEPA,0903,136;%%
  
  %\cite{Jackiw:1981ee}
\bibitem{Jackiw:1981ee}
  R.~Jackiw and P.~Rossi,
  %``Zero Modes Of The Vortex - Fermion System,''
  Nucl.\ Phys.\  B {\bf 190} (1981) 681.
  %%CITATION = NUPHA,B190,681;%%
  
%\cite{Witten:1984eb}
\bibitem{Witten:1984eb}
  E.~Witten,
  %``Superconducting Strings,''
  Nucl.\ Phys.\  B {\bf 249} (1985) 557.
  %%CITATION = NUPHA,B249,557;%%
  
%\cite{Narayanan:1994gw}
\bibitem{Narayanan:1994gw}
  R.~Narayanan and H.~Neuberger,
  %``A Construction of lattice chiral gauge theories,''
  Nucl.\ Phys.\  B {\bf 443} (1995) 305
  [arXiv:hep-th/9411108].
  %%CITATION = NUPHA,B443,305;%%
  %\cite{RandjbarDaemi:1995cq}
%\bibitem{RandjbarDaemi:1995cq}
  S.~Randjbar-Daemi and J.~A.~Strathdee,
  %``Chiral Fermions On The Lattice,''
  Nucl.\ Phys.\  B {\bf 443} (1995) 386
  [arXiv:hep-lat/9501027].
  %%CITATION = NUPHA,B443,386;%%

%\cite{RandjbarDaemi:1994tj}
\bibitem{RandjbarDaemi:1994tj}
  S.~Randjbar-Daemi and J.~A.~Strathdee,
  %``Gravitational Lorentz Anomaly From The Overlap Formula In Two-Dimensions,''
  Phys.\ Rev.\  D {\bf 51} (1995) 6617
  [arXiv:hep-th/9501012].
  %%CITATION = PHRVA,D51,6617;%%
  
  %\cite{Rubakov:1983bb}
\bibitem{Rubakov:1983bb}
  V.~A.~Rubakov and M.~E.~Shaposhnikov,
  %``Do We Live Inside A Domain Wall?,''
  Phys.\ Lett.\  B {\bf 125} (1983) 136.
  %%CITATION = PHLTA,B125,136;%%
%\cite{Rubakov:1983bz}
%\bibitem{Rubakov:1983bz}
  V.~A.~Rubakov and M.~E.~Shaposhnikov,
  %``Extra Space-Time Dimensions: Towards A Solution To The Cosmological
  %Constant Problem,''
  Phys.\ Lett.\  B {\bf 125} (1983) 139.
  %%CITATION = PHLTA,B125,139;%%



%\cite{Sundrum:1998sj}
\bibitem{Sundrum:1998sj}
  R.~Sundrum,
  %``Effective field theory for a three-brane universe,''
  Phys.\ Rev.\  D {\bf 59} (1999) 085009
  [arXiv:hep-ph/9805471].
 %%CITATION = PHRVA,D59,085009;%%



%\cite{Carroll:2003db}
\bibitem{Carroll:2003db}
  S.~M.~Carroll and M.~M.~Guica,
  %``Sidestepping the cosmological constant with football-shaped extra
  %dimensions,''
  arXiv:hep-th/0302067.
  %%CITATION = HEP-TH 0302067;%%
  I.~Navarro,
  %``Codimension two compactifications and the cosmological constant  problem,''
  JCAP {\bf 0309}, 004 (2003)
  [arXiv:hep-th/0302129].
  %%CITATION = HEP-TH 0302129;%%





\bibitem{nonsusysphere}
S.~Randjbar-Daemi, A.~Salam and J.~A.~Strathdee, %``Spontaneous
%Compactification In Six-Dimensional Einstein-Maxwell Theory,''
Nucl.\ Phys.\ B {\bf 214} (1983) 491.



\bibitem{Nishino:1984gk}
  H.~Nishino and E.~Sezgin,
  %``Matter And Gauge Couplings Of N=2 Supergravity In Six-Dimensions,''
  Phys.\ Lett.\ B {\bf 144} (1984) 187.
  %%CITATION = PHLTA,B144,187;%%
  %\cite{Randjbar-Daemi:1985wc}
%\cite{Nishino:1986dc}

%\cite{Randjbar-Daemi:1985wc}
\bibitem{seifanomaly}
  S.~Randjbar-Daemi, A.~Salam, E.~Sezgin and J.~A.~Strathdee,
 %``An Anomaly Free Model In Six-Dimensions,''
  Phys.\ Lett.\ B {\bf 151} (1985) 351.
  %%CITATION = PHLTA,B151,351;%%
  %\cite{RandjbarDaemi:2004qr}
%\bibitem{RandjbarDaemi:2004qr}
  S.~Randjbar-Daemi and E.~Sezgin,
  %``Scalar potential and dyonic strings in 6d gauged supergravity,''
  Nucl.\ Phys.\  B {\bf 692} (2004) 346
  [arXiv:hep-th/0402217].
  %%CITATION = NUPHA,B692,346;%%



%\cite{Avramis:2005qt}
\bibitem{Avramis:2005qt}
  S.~D.~Avramis, A.~Kehagias and S.~Randjbar-Daemi,
 % ``A new anomaly-free gauged supergravity in six dimensions,''
  JHEP {\bf 0505} (2005) 057
  [arXiv:hep-th/0504033].
  %%CITATION = HEP-TH 0504033;%%

%\cite{Avramis:2005hc}
\bibitem{Avramis:2005hc}
  S.~D.~Avramis and A.~Kehagias,
 % ``A systematic search for anomaly-free supergravities in six dimensions,''
  JHEP {\bf 0510} (2005) 052
  [arXiv:hep-th/0508172].
  %%CITATION = HEP-TH 0508172;%%

%\cite{Salvio:2007mb}

%\cite{Suzuki:2005vu}
\bibitem{Suzuki:2005vu}
  %\cite{Suzuki:2005vu}
%\bibitem{Suzuki:2005vu}
  R.~Suzuki and Y.~Tachikawa,
%  ``More anomaly-free models of six-dimensional gauged supergravity,''
  J.\ Math.\ Phys.\  {\bf 47}, 062302 (2006)
  [arXiv:hep-th/0512019].
  %%CITATION = JMAPA,47,062302;%%



%\cite{Cvetic:2003xr}
\bibitem{Cvetic:2003xr}
  M.~Cvetic, G.~W.~Gibbons and C.~N.~Pope,
  %``A string and M-theory origin for the Salam-Sezgin model,''
  Nucl.\ Phys.\  B {\bf 677} (2004) 164
  [arXiv:hep-th/0308026].
  %%CITATION = NUPHA,B677,164;%%


%\cite{Gibbons:2003di}
\bibitem{Gibbons:2003di}
  G.~W.~Gibbons, R.~Guven and C.~N.~Pope,
  %``3-branes and uniqueness of the Salam-Sezgin vacuum,''
  Phys.\ Lett.\  B {\bf 595} (2004) 498
  [arXiv:hep-th/0307238].
  %%CITATION = PHLTA,B595,498;%%

  %\cite{Salam:1984cj}
\bibitem{salamsezgin}
  A.~Salam and E.~Sezgin,
  %``Chiral Compactification On Minkowski X S**2 Of N=2 Einstein-Maxwell
  %Supergravity In Six-Dimensions,''
  Phys.\ Lett.\  B {\bf 147} (1984) 47.
  %%CITATION = PHLTA,B147,47;%%


%\cite{Aghababaie:2003ar}
\bibitem{Aghababaie:2003ar}
  Y.~Aghababaie {\it et al.},
  %``Warped brane worlds in six dimensional supergravity,''
  JHEP {\bf 0309} (2003) 037
  [arXiv:hep-th/0308064].
  %%CITATION = JHEPA,0309,037;%%


%\cite{Burgess:2006ds}
\bibitem{kicking}
%\cite{Burgess:2006ds}
%\bibitem{Burgess:2006ds}
  C.~P.~Burgess, C.~de Rham, D.~Hoover, D.~Mason and A.~J.~Tolley,
  %``Kicking the rugby ball: Perturbations of 6D gauged chiral supergravity,''
  JCAP {\bf 0702} (2007) 009
  [arXiv:hep-th/0610078].
  %%CITATION = JCAPA,0702,009;%%






%\cite{Salvio:2007mb}
\bibitem{Salvio:2007mb}
  A.~Salvio,
  %``Aspects of physics with two extra dimensions,''
  arXiv:hep-th/0701020.
  %%CITATION = HEP-TH/0701020;%%

%\cite{Lee:2007ib}
\bibitem{gravitino}
  H.~M.~Lee and A.~Papazoglou,
  %``Gravitino in six-dimensional warped supergravity,''
  Nucl.\ Phys.\  B {\bf 792} (2008) 166
  [arXiv:0705.1157 [hep-th]].
  %%CITATION = NUPHA,B792,166;%%

%\cite{Salvio:2009mp}
\bibitem{Salvio:2009mp}
  A.~Salvio,
  %``Brane Gravitational Interactions from 6D Supergravity,''
  Phys.\ Lett.\  B {\bf 681} (2009) 166
  [arXiv:0909.0023 [hep-th]].
  %%CITATION = PHLTA,B681,166;%%



%\cite{RandjbarDaemi:2002pq}
\bibitem{seifgen}
  S.~Randjbar-Daemi and M.~Shaposhnikov,
  %``A formalism to analyze the spectrum of brane world scenarios,''
  Nucl.\ Phys.\  B {\bf 645} (2002) 188
  [arXiv:hep-th/0206016].
  %%CITATION = NUPHA,B645,188;%%

%\cite{Nicolai:1984jg}
\bibitem{Nicolai:1984jg}
  H.~Nicolai and C.~Wetterich,
  %``On The Spectrum Of Kaluza-Klein Theories With Noncompact Internal Spaces,''
  Phys.\ Lett.\ B {\bf 150} (1985) 347.
  %%CITATION = PHLTA,B150,347;%%

%\cite{Gibbons:1986wg}
\bibitem{Gibbons:1986wg}
  G.~W.~Gibbons and D.~L.~Wiltshire,
 % ``Space-Time As A Membrane In Higher Dimensions,''
  Nucl.\ Phys.\ B {\bf 287}, 717 (1987)
  [arXiv:hep-th/0109093].
  %%CITATION = HEP-TH 0109093;%%

%\cite{RandjbarDaemi:2006gf}
\bibitem{RandjbarDaemi:2006gf}
  S.~Randjbar-Daemi, A.~Salvio and M.~Shaposhnikov,
  %``On the decoupling of heavy modes in Kaluza-Klein theories,''
  Nucl.\ Phys.\  B {\bf 741} (2006) 236
  [arXiv:hep-th/0601066].
  %%CITATION = NUPHA,B741,236;%%
%\cite{Salvio:2006mh}
%\bibitem{Salvio:2006mh}
  A.~Salvio,
  %``4D effective theory and geometrical approach,''
  AIP Conf.\ Proc.\  {\bf 881} (2007) 58
  [arXiv:hep-th/0609050].
  %%CITATION = APCPC,881,58;%%

%\cite{Schwindt:2003er}
\bibitem{Schwindt:2003er}
  J.~M.~Schwindt and C.~Wetterich,
%  ``Holographic branes,''
  Phys.\ Lett.\ B {\bf 578} (2004) 409
  [arXiv:hep-th/0309065].
  %%CITATION = HEP-TH 0309065;%%


\bibitem{seifinstabilities}
%  %%CITATION = NUPHA,B214,491;%%
  S.~Randjbar-Daemi, A.~Salam and J.~A.~Strathdee,
  %``Instability Of Higher Dimensional Yang-Mills Systems,''
  Phys.\ Lett.\ B {\bf 124} (1983) 345
  [Erratum-ibid.\ B {\bf 144} (1984) 455].
  %%CITATION = PHLTA,B124,345;%%
%\cite{Dvali:2001qr}
%\bibitem{Dvali:2001qr}
  G.~R.~Dvali, S.~Randjbar-Daemi and R.~Tabbash,
  %``The origin of spontaneous symmetry breaking in theories with large  extra
  %dimensions,''
  Phys.\ Rev.\  D {\bf 65} (2002) 064021
  [arXiv:hep-ph/0102307].
  %%CITATION = PHRVA,D65,064021;%%


%\cite{Burgess:2008az}
\bibitem{endpoint}
  C.~P.~Burgess, S.~L.~Parameswaran and I.~Zavala,
  %``The Fate of Unstable Gauge Flux Compactifications,''
  arXiv:0812.3902 [hep-th].
  %%CITATION = ARXIV:0812.3902;%%



\end{thebibliography}
\end{document}